\documentclass[aps12,pre,twocolumn,showpacs,preprintnumbers,amsmath,amssymb, amsthm, floatfix,superscriptaddress]{revtex4}

\usepackage{appendix}
\usepackage[usenames]{color}
\usepackage[dvips]{graphicx}
\usepackage{amsmath,amssymb,amsthm}
\usepackage[mathscr]{eucal}

\newcommand{\be}{\begin{eqnarray*}}
\newcommand{\ee}{\end{eqnarray*}}
\usepackage{verbatim}
\renewcommand{\vec}{\mathbf}
\def\Tr#1{{\rm Tr}\left( #1 \right)}
\def\Det#1{{\rm Det}\!\left( #1 \right)}

\def\matr#1{\underline{\underline{{#1}}}}

\def\dm{ \matr{\delta} }
\def\n{{\hat{\bf{n} }}}

\def\half{{\textstyle \frac{1}{2}}}

\newcommand{\degree}{^{\rm o}}

\usepackage{hyperref}
\usepackage{epsfig,graphicx}% Include figure files
\usepackage{dcolumn}% Align table columns on decimal point
\usepackage{bm}% bold math

\begin{document}

\title{Elasticity of Polydomain Liquid Crystal Elastomers}
\author{J. S. Biggins\email{john.biggins@gmail.com}}
\author{M. Warner}
\affiliation{Cavendish Laboratory, University of Cambridge, Cambridge, CB3 0HE, United Kingdom}
\author{K. Bhattacharya}
\affiliation{Division of Engineering and Applied Science, California Institute of Technology}

\date{\today}
\begin{abstract}
We model polydomain liquid-crystal elastomers by extending the neo-classical soft and semi-soft free energies used successfully to describe monodomain samples. We show that there is a significant difference between polydomains cross-linked in homogeneous  high symmetry  states then cooled to low symmetry polydomain states and those cross-linked directly in the low symmetry polydomain state. For example, elastomers cross-linked in the isotropic state then cooled to a nematic polydomain will, in the ideal limit, be perfectly soft, and with the introduction of non-ideality, will deform at very low stress until they are macroscopically aligned. The director patterns observed in them will be disordered, characteristic of combinations of random deformations, and  not disclination patterns. We expect these samples to exhibit elasticity significantly softer than monodomain samples. Polydomains cross-linked in the nematic polydomain state will be mechanically harder and contain characteristic schlieren director patterns. The models we use for polydomain elastomers are spatially heterogeneous, so rather than solving them exactly we elucidate this behavior by bounding the energies using Taylor-like (compatible test strain fields) and Sachs (constant stress) limits extended to non-linear elasticity.  Good agreement is found with experiments that reveal the supersoft response of some polydomains.  We also analyze smectic polydomain elastomers and propose that polydomain SmC* elastomers cross-linked in the SmA monodomain state are promising candidates for low field electrical actuation.
\end{abstract}
\maketitle

\section{Introduction}\setcounter{equation}{0}
Liquid crystal elastomers (LCE's) are rubbery networks of entropically dominated polymer chains that exhibit mobile liquid crystalline order. The elasticity of liquid crystal elastomers is a rich and subtle subject because the liquid crystal order is anisotropic, the strains involved are often large and the elastomer responds very softly to some stresses but more conventionally to others. Monodomain liquid crystal elastomers (elastomers in which the liquid crystal director does not vary from point to point in the sample) were first successfully synthesized in 1991 \cite{Kupfer1991}. Since then many remarkable phenomena have been observed in nematic elastomers --- extremely soft elasticity and textured deformations \cite{ssplateau}, dramatic spontaneous deformations at the transition between the isotropic and aligned states \cite{clarke01}, rotation of the nematic director by the application of stress \cite{semisoftrot} or electric fields \cite{urayama06}.  Elongations go back to molecular shape depending on the magnitude of nematic order.  Mechanically soft response depends on rotating the director with concomitant  shears and elongations that accommodate the rotating long direction of the chain distribution \cite{soft1,soft2,verwey96}, summarised in \cite{LCE}.
 The very special deformations that accompany soft response of monodomains depend critically on the direction of the initial director to the overall extension direction.  It therefore seems remarkable that polydomain elastomers are also capable of soft response to macroscopic deformations \cite{Hotta01,Uryamapoly} that also progressively align the directors.  One would expect neighbouring regions with initially different directors and therefore differing soft strains would suffer mechanical incompatibility.  Differing strategies for overcoming these apparent geometric impediments to soft response have been proposed \cite{terentjevrandom,terentjevpoly,Uchida:99a,Uchida:99b,Uchida:00,Biggins:09}; we discuss this work in detail later.
 
 In this paper we  attack the central geometric issues of how extremely soft response is attained in isotropic genesis polydomain elastomers and why nematic genesis elastomers are harder.  We bound the effect of semi-softness in the former response.    Recent experiments \cite{Uryamapoly} show remarkable, genesis-dependent, super-softness of polydomain nematic elastomers.  The different response of elastomers of differing genesis is a result we share with Uchida, though our route to geometric compatibility is rather different.  These results may be significant to the development of LCE actuators since polydomain elastomers are significantly easier to synthesize than their monodomain counterparts, can exert much larger forces because they are not limited to the thin film geometries needed for making monodomains, and can respond to external stimuli, such as polarised light,  without having such large internal (mechanical) fields to overcome.

\subsection{Models of Liquid Crystal Elastomers}
\begin{figure}
\includegraphics[width=3cm]{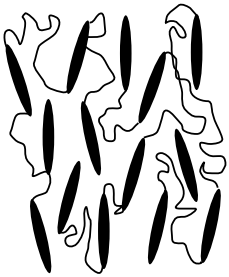}
\caption{Diagram of a nematic polymer melt. The black lines represent polymer strands and the ellipsoids are nematic rods.}\label{fig:polymermelt}
\end{figure}

\begin{figure}
\includegraphics[width=8cm]{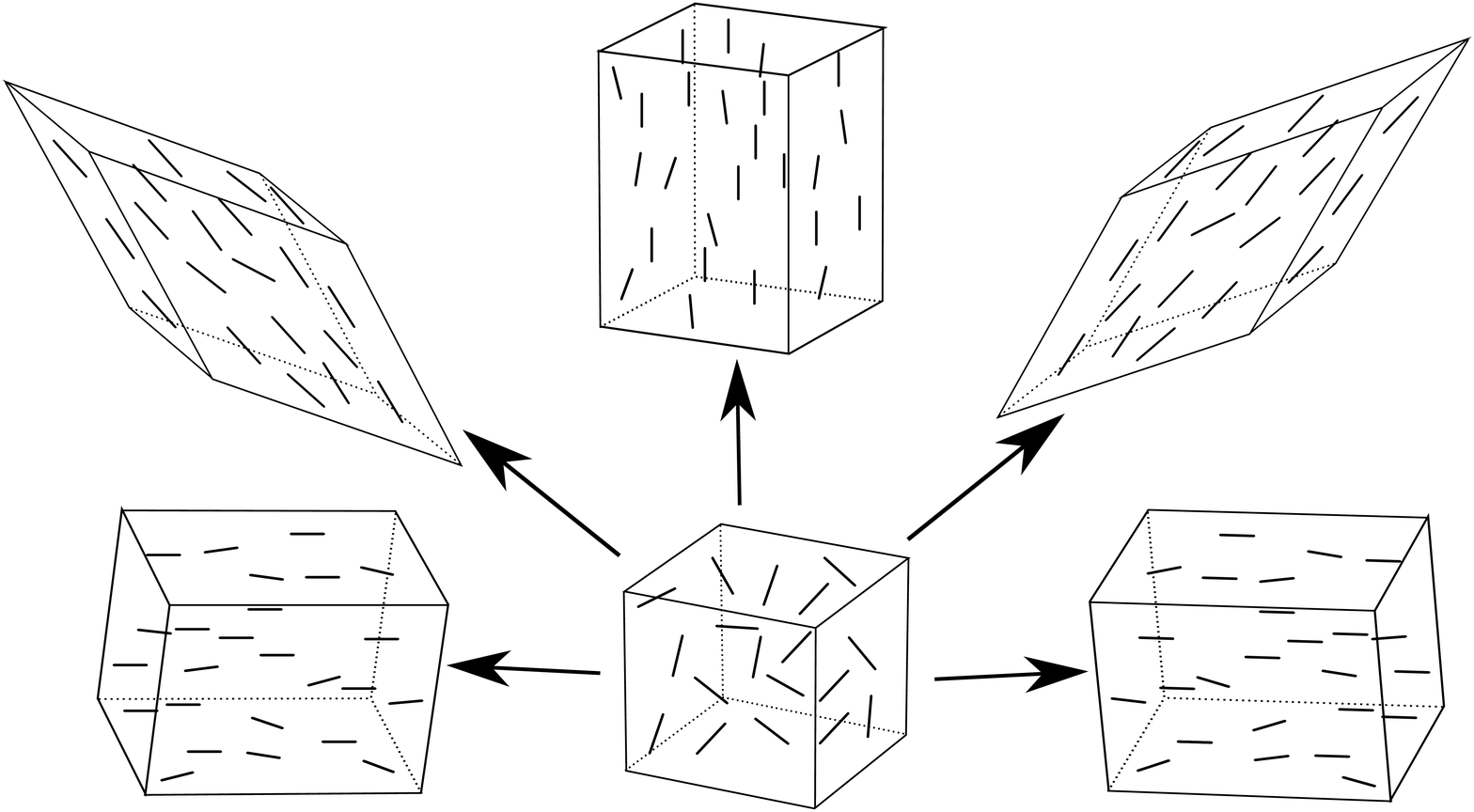}
\caption{A liquid crystalline polymer is cross-linked in the high temperature isotropic state - centre, bottom row On cooling, a nematic phase is formed and the elastomer stretches along the new nematic director. Since any director could have been chosen there are many equivalent low energy states, shown in an arc around the isotropic state. Deformations around the arc will not cost energy.}\label{fig:transitionstretch}
\end{figure}

Microscopically liquid crystal elastomers can be thought of as the result of cross-linking a liquid-crystalline polymer melt. The melt consists of long writhing polymer molecules that incorporate many short rod-like units - fig.\ \ref{fig:polymermelt}.
Below a transition temperature the rod like mesogens align to form a liquid crystal phase. The introduction of cross-links between the polymer chains prevents them from flowing past each other making the material a solid elastomer rather than a liquid melt. However, sufficiently low density cross-linking does not stop the extensive thermal motion of the chains or prevent the formation of liquid crystal order, so the resulting solid is a liquid crystalline elastomer. The novel elastic properties of a liquid crystal elastomer can most easily be understood by considering an elastomer that was cross-linked in the high temperature isotropic state then cooled to a nematic monodomain in which the nematic director is constant throughout the sample. After the nematic phase has formed the polymers are more likely to adopt conformations which are extended along the nematic director, so the elastomer undergoes a uniaxial stretch along the nematic director - fig.\ \ref{fig:transitionstretch}.
However, since the high temperature cross-linking state is completely isotropic, any direction could have been chosen for the nematic director so there is a degenerate set of low energy states each with a different director and a different deformation with respect to the cross-linking state. Deformations that take the elastomer from one low energy state to another must not cost any energy to impose and result in rotation of the liquid crystal director through the elastomer. This symmetry-driven phenomenological explanation of soft elastic modes was first presented by Golubovic and Lubensky \cite{GL}; see \cite{LCE} for the theorem's application to liquid crystal elastomers.

Experimentally elastomers with spontaneous elongations at the isotropic-nematic transition by factors of 4 \cite{clarke01} have been synthesized, so it is clear that any theory of liquid crystal elastomers must be fully non-linear in the strain tensor. Moreover, strains induce director rotations of up to 90$\degree$, hence finite rotation theory is also needed. The simplest model that we can write down for the system discussed above is simply a neo-Hookean free energy that incorporates a large spontaneous deformation
\begin{equation}
F=\half \mu \Tr{ \matr{\gamma}^T\cdot\matr{\ell}^{-1}\cdot\matr{\gamma}}\label{monoenergy}
\end{equation}
where $\mu$ is a shear modulus, $\matr{\gamma}$ is the deformation gradient from the  cross-linking state (here the isotropic state) and, if $\dm$ is the identity matrix and $r$ is a constant material parameter, $\matr{\ell}$ is derived from the final state liquid crystal director $\n$ as $\matr{\ell}=r^{-1/3}(\dm+(r-1)\n\n)$ (a notation differing from \cite{LCE} by a factor of $r^{_1/3}$).  In the isotropic phase $r=1$ and this form reduces to a standard neo-Hookean energy, but on entering the nematic phase $r>1$ and this energy is minimized by any choice of $\vec{n}$ provided it is accompanied by a spontaneous deformation $\matr{\gamma}=\matr{\ell}^{1/2}$ which is an elongation from the isotropic state along $\n$ by a factor of $r^{1/3}$. Elastomers that suffer spontaneous deformations by factors between 1.05 and 4 have been observed, implying values for $r$ between 1 and 60 \cite{clarke01}. As for all elastomers, LCE's are essentially incompressible so we require that $\Det{\matr{\gamma}}=1$, a condition satisfied for example by spontaneous deformations, $\Det{\matr{\ell}^{1/2}} = 1$. We can rewrite this energy in terms of deformations from a minimum energy, relaxed nematic  state with director $\vec{n}_0$ by substituting $\matr{\gamma_n}=\matr{\gamma}\cdot\matr{\ell_0}^{-1/2}$ to give
\begin{equation}
F=\half \mu \Tr{ \matr{\ell_0}\cdot\matr{\gamma_n}^T\cdot\matr{\ell}^{-1}\cdot\matr{\gamma_n}},
\end{equation}
where $\vec{n}_0$ is the nematic director in the reference state and $\vec{n}$ is the director in the final state. A microscopic derivation of these free energies is given in \cite{trformula, LCE} which is a simple extension of the classical theory of rubber based on Gaussian polymer chains. The microscopic model provides one key additional insight --- in the idealized limit of compositionally homogeneous Gaussian chains the local energy does not remember the cross-linking state, so the above energy for an elastomer cross-linked in the isotropic state then cooled to the nematic state is also appropriate if the elastomer is cross-linked in the nematic state with director $\vec{n}_0$. Real chains are never perfectly Gaussian or completely compositionally homogeneous, so real elastomers cross-linked in the nematic state do remember the director at cross-linking and their free energy is minimized when their director aligns in this direction. This effect is modeled by adding a ``non-ideal'' term to the energy which favors alignment of the director along $\vec{n_0}$ giving a free energy density such as
 \begin{equation}
F=\half \mu \Tr{\matr{\gamma}\cdot\matr{\ell_0}\cdot\matr{\gamma}^T\cdot\matr{\ell}^{-1}+\alpha\matr{\gamma}\cdot\left(\dm-\n_0\n_0\right)\cdot\matr{\gamma}^T\cdot\n\n},
\end{equation}
the most general form at order $\matr{\lambda}^2$ \cite{bigginssemisoft}. The coefficient $\alpha$ determines how far the elastomer deviates from ideally soft behavior.

Liquid crystal elastomers can also be prepared in layered (smectic) liquid crystal phases. Elastomers have been synthesized in both SmA phases in which the director is along the layer normal and SmC phases where there is a preferred tilt angle $\theta$ between the director and the layer normal. Smectic elastomers are modeled by assuming that the layers deform affinely and that there are energy penalties for deformations that change the layer spacing or cause the director to rotate away from its preferred tilt angle. This means that SmA elastomers do not have any soft elastic modes since the layers deform affinely and rotation of the director away from the layer normal costs energy. However, the elasticity of SmA elastomers is still a complicated subject because the modulus for stretching the inter-layer spacing is much higher than the modulus for in-plane stretches so their elasticity is essentially two dimensional. SmC elastomers in which the director makes a fixed angle with the layer normal do still have soft modes because the director can rotate in a cone around the layer normal without changing the inter-layer spacing or deviating from the preferred tilt angle.  Accordingly we shall predict unusual mechanical response for polydomain SmC elastomers.

\subsection{Polydomain and Monodomain Elastomers}
An LCE is a monodomain if the director is the same at every point in the relaxed elastomer and a polydomain if it points in different directions at different points in the relaxed elastomer. Synthesizing monodomains is difficult because, in order to make the elastomer choose the same director at each point, this direction must be imprinted on the elastomer at cross-linking, normally by cross-linking under uniaxial stress. If the elastomer is cross-linked without any such imprinting, in either the isotropic or the nematic state, then it forms a polydomain. Polydomains and monodomains are easily distinguished because monodomains are highly transparent and exhibit large spontaneous deformations at the isotropic-nematic transition while polydomains have no macroscopic spontaneous deformation, and are opaque in the nematic state (because the gradients in the director scatter light) but become transparent if they are stretched enough to align the director throughout the sample \cite{Clarke98}.

We extend the monodomain energies in the previous section to polydomain energies by allowing $\n$, $\n_0$ and $\matr{\gamma}$ to become spatially varying fields. We define $\matr{\gamma}(\vec{x})= \nabla \vec{y}$ to be the deformation gradient from the cross-linking state. This means that, ignoring non-ideal terms, the energy function for a polydomain cross-linked in the nematic state will be
 \begin{equation}
F=\half \mu \Tr{\matr{\gamma}\cdot\matr{\ell_0}\cdot\matr{\gamma}^T\cdot\matr{\ell}^{-1}}\end{equation}
where $\matr{\ell_0}$ is derived from $\n_0(\vec{x})$ which is the nematic director at cross-linking. In this case the form of this field will be the nematic disclination pattern  present in the nematic-melt before cross-linking. This energy is very different to the monodomain energy from which it was derived because it has a very significant $\vec{x}$ dependence in $\n_0(\vec{x})$. However, the energy function for a polydomain cross-linked in the isotropic state is simply
\begin{equation}
F=\half \mu \Tr{ \matr{\gamma}^T\cdot\matr{\ell}^{-1}\cdot\matr{\gamma}}
\end{equation}
which is exactly the same as the corresponding monodomain energy, eqn.\ \eqref{monoenergy}. There is no intrinsic spatial variation in this function, which is not surprising since the cross-linking state is completely isotropic and homogeneous. However, monodomains and polydomains cross-linked in the isotropic state are manifestly different. We propose that this is because of the form of the non ideal terms that must be added to this energy. In the monodomain case the elastomer is cross-linked in the presence of a uniaxial stress which imprints a constant preferred direction $\vec{n_0}$ (in effect a field) on the elastomer so that when it cools to the nematic state it adopts the same nematic director everywhere, leading us to a mono-domain energy
\begin{align}
F=\half \mu \Tr{\matr{\gamma}\cdot\matr{\gamma}^T\cdot\matr{\ell}^{-1}+\alpha r^{1/3}\matr{\gamma}\cdot\left(\dm-\n_0\n_0\right)\cdot\matr{\gamma}^T\cdot\n\n}.
\end{align}
In the polydomain case there is no imprinting so there is very little to break the isotropy of the cross-linking state and introduce any non-ideal terms at all. However, weak mechanisms do exist such as cross-linking molecules having a rod like character which impose an additional direction locally on the network \cite{vwmodel2}.  Consequently, in any finite region the cross-linking rods develop a slight average orientation \cite{terentjevpoly,ClarkePM}. These mechanisms permit the inclusion of a very small non-ideal term with a spatially varying preferred direction $\n_0(\vec{x})$. It is this distinction between the large homogeneous non-ideal term in the monodomain case and the small spatially varying term in the polydomain case that drives the distinction between the two systems.

\subsection{Polydomains - Macroscopically Hard or Soft?}
In this paper we argue that whether polydomain elastomers are macroscopically hard or soft depends on the relative symmetry of the cross-linking state and the final polydomain state. In this section we will give a qualitative overview of the cause of this behaviour, then in subsequent sections we will work through three examples of possible polydomain systems.

\begin{figure}
\includegraphics[width=8cm]{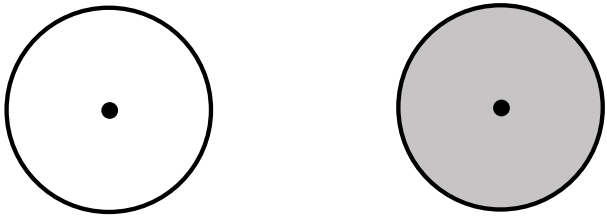}
\caption{Left - the ring of deformations leading to low energy states for a liquid crystal elastomer. The dot in the center is the high symmetry parent state from which the broken-symmetry, low energy states on the ring are derived. Right - visualization of the quasi-convex hull of the set of low energy deformations. These deformations, inside the ring (shaded grey), can be imposed with minimal energy because they can be constructed out of textures of the soft deformations that each lie on different points of the ring.}\label{fig:ring}
\end{figure}

An ideal liquid crystal elastomer energy typically locally has a set of deformations that minimize the free-energy. These are generated by a symmetry breaking phase transition from a high temperature parent state that is accompanied by a deformation. Since the transition breaks a symmetry there are many ground states, each with a different deformation with respect to the parent state. We visualize this set of energy minimizing deformations as a ring with the parent state at the centre --- fig.\ \ref{fig:ring}. Such a system is vulnerable to the formation of textured deformations since, if a deformation on the interior of the ring is imposed this is not a low energy state, but it is possible that the energy of the deformation can be reduced to zero by the elastomer splitting into many small regions each of which undergoes one of the low energy deformations in such a way that the macroscopic deformation is what was imposed. The ability of LCE's to form such textured deformations has been the subject of several studies --- \cite{semisoftrot, DeDo2002, Contisemisoft, contiadams}. We represent the set of all deformations that can minimize the energy after the formation of the most advantageous textured deformations, the quasi-convex hull (QCH) of the set of energy minimizing deformations, as the interior of the ring of energy minimizing deformations --- fig.\ \ref{fig:ring}.

A polydomain sample cross-linked in the high symmetry (high temperature) state is effectively cross-linked at the centre of the quasi-convex hull and, on cooling to the aligned state, forms the \textit{same} quasi-convex hull at each point. Although at each point the elastomer has undergone a deformation that takes it to the boundary of the set, these deformations are put together in an elastically compatible way  so that the elastomer as a whole is in a textured state at the centre of the set. Such an elastomer can be deformed macroscopically softly simply by moving the constituent domains around the quasi-convex hull, which will cause different soft textures to evolve as the elastomer as a whole moves across the quasi-convex hull.

\begin{figure}
\includegraphics[width=4cm]{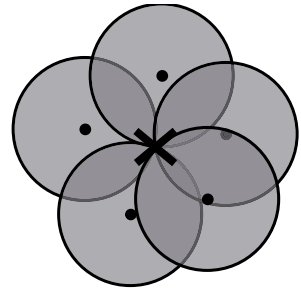}
\caption{If the elastomer is cross-linked in the low symmetry polydomain state then, although every domain still has a quasi-convex hull of the same form, each domain is cross-linked at the boundary of its quasi-convex hull and requires a different deformation to get to the its centre. The quasi-convex hulls of a few domains are illustrated in the diagram, and the cross represents the cross-linking configuration. The cross is the only point in every quasi-convex hull so if the elastomer deforms away from this point some domains will leave their convex hull and the response will be hard.}\label{fig:nemhull}
\end{figure}

In contrast, if the elastomer is cross-linked in the low temperature, symmetry-broken state then,  although each point still has a quasi-convex hull of the same form each domain sits on the edge of its quasi-convex hull and the deformation required to take each domain back to the center of its hull is different. This corresponds to a picture like fig.\ \ref{fig:nemhull}. Although each domain has a hull and soft modes, there are no deformations that are soft for all domains so the elastomer is macroscopically hard.
As discussed in the previous section the distinction between a polydomain cross-linked in the high symmetry state and  a monodomain is driven by the addition of different non ideal terms. This is easily visualized in terms of the above sets - the ideal monodomain and high symmetry cross-linking polydomain have the same quasi-convex hull at each point, but the addition of non-ideal terms breaks the degeneracy of the states in the quasi-convex hull, making one a unique global minimum. In the monodomain case this is a point on the boundary of the set with a uniform director throughout the sample. In the polydomain case, the non-ideal terms  vary spatially such that  a different low energy state on the boundary of the QCH is favoured for different points. The global minimum is the point at the centre of the set, no deformation, and is achieved by a complicated textured, low energy, deformation of the different regions of the polydomain.

The above analysis of soft  polydomains has a very simple physical interpretation. If an elastomer is cross-linked in the high symmetry state then cooled to the low symmetry state there is no energetic penalty (except small deviations from ideality) to stop it breaking the symmetry in the same way at every point in the elastomer.
 Therefore the macroscopic deformations that take the polydomains into these well-aligned states must be soft.  This will not be true if the crosslinking is in the low symmetry polydomain state.

In the remainder of this paper we will analyze nematic polydomains cross-linked in the isotropic state and the nematic state and SmC polydomains cross-linked in the SmA monodomain state. The first and last of these are examples of soft polydomains, while the middle is a hard polydomain.

\section{Nematic Polydomains}
\subsection{Formulating the elasticity problems}
Following the discussion in the introduction, we expect a profound difference between nematic polydomains cross-linked in the isotropic and nematic state. We further distinguish between ideal and non-ideal polydomains.  Ideal systems have no locally preferred director orientation and therefore locally completely soft modes. The addition of non-ideal terms slightly favours a particular local director orientation making the modes that were previously completely soft now cost a small amount of energy. This leads us to four types of nematic polydomain corresponding to four different free energies,
\begin{equation}
F=\begin{cases}
\half\mu\Tr{\matr{\gamma}\cdot\matr{\gamma}^T\cdot{\matr{\ell}^{-1}}} &\text{iI}\\
\half\mu\Tr{\matr{\gamma}\cdot\matr{\gamma}^T\cdot{\matr{\ell}^{-1}}+\alpha r^{1/3} \matr{\gamma}\left(\dm-\n_0\n_0\right)\matr{\gamma}^T\n\n}  & \text{nI}\\
\half\mu\Tr{\matr{\gamma}\cdot\matr{\ell}_0\cdot\matr{\gamma}^T\cdot{\matr{\ell}^{-1}}} & \text{iN}\\
\half\mu\Tr{\matr{\gamma}\cdot\matr{\ell}_0\cdot\matr{\gamma}^T\cdot{\matr{\ell}^{-1}}+\alpha \matr{\gamma}\left(\dm-\n_0\n_0\right)\matr{\gamma}^T\n\n}  & \text{nN}\\
\end{cases}
\end{equation}
where iI/nI denote  ideal/non-ideal elastomers crosslinked in the isotropic state, and iN/nN  ideal/non-ideal elastomers crosslinked in the  nematic state. The bulk modulus of elastomers is several orders of magnitude higher than the shear modulus, so all deformations are volume preserving, that is $\Det{\matr{\gamma}}=1$.  The preferred direction $\n_0$ is discussed below.

Having identified four types of polydomain, ideal and non-ideal, with either a nematic or an isotropic crosslinking state (genesis), we wish to study the energetic cost of imposing macroscopically homogeneous stretches on large blocks of these different types of polydomains. By large, we mean large enough to contain very many domains and be macroscopically isotropic. In each case we take the cross-linking state of the elastomer as the reference configuration from which deformations will be measured, and define the displacement field from this state as $\vec{y}(\vec{x})$, so that the local deformation gradient is $\matr{\gamma}(\vec{x})=\nabla\vec{y}$. We then wish to study the energy of a large sample occupying a domain $\Omega$ (with boundary $\delta\Omega$) in the reference configuration, that is subject to a macroscopic deformation $\matr{\lambda}$ after it has adopted the most favourable internal deformation and director pattern. We define this relaxed energy function as
\begin{equation}
F^{r}\!(\matr{\lambda}, \n_0(\vec{x}))=\!\!\min_{ \substack{\vec{y}(\vec{x}) \, \rm{s.t.} \\ \vec{y} = \matr{\lambda} \cdot \vec{x} \\ {\rm on\ } \delta\Omega}}  \min_{\n{(\vec{x})}}  \frac{1}{\rm{Vol} \, \Omega}\!\int\!\! F(\nabla{ \vec{y} },\vec{n}(\vec{x}),\vec{n}_0(\vec{x})) d{\vec{x}},
\end{equation}
and the four different types of polydomain correspond to four different choices for $ F(\nabla{y},\vec{n}(\vec{x}),\vec{n}_0(\vec{x}))$.

There is an important distinction between the two fields $\n_0(\vec{x})$, which defines the local preferred nematic alignment, and $\n(\vec{x})$ which is the nematic field after deformation. The former is a fixed field for a given polydomain that encodes all its spatial heterogeneity, while the latter is a variable field which the elastomer will adjust to minimize its free energy as it evolves under the macroscopic $\matr{\lambda}$. Furthermore, in the nematic genesis case we expect the form of $\n_0(\vec{x})$ to be a disclination texture, while  in the isotropic genesis case it will be a random field arising from the sources of disorder in the crosslinking state.  In the latter case $\vec{n}_0$ does not correspond to the equilibrium director pattern at zero stress, but rather locally it is the director that a domain would adopt if it were unconstrained by its neighbors (e.g.\ by cutting it out of the sample). These locally optimal strain fields $\matr{\gamma}(\vec{x})$ associated with this director pattern are extremely unlikely to be  compatible deformations (be the gradient of a continuous displacement field) and thus would require the sample to fracture.

Our task is to find the lowest energy, compatible strain fields -- in general a difficult task.  It is not clear that they are very subtle fields associated with a continuous director variation (but still compatible) or are compatible combinations (textures and laminates) of strains as are observed in the deformation of monodomains under boundary constraints at variance with soft deformations \cite{semisoftrot}.  We shall proceed to find bounds on the elastic energy.  In the ideal case, we know that textures of deformations each on the boundary of the QCH can give zero energy cost for macroscopic deformations within the QCH and thus give an exact value for the energy (a continuous field could not do better than this choice of test field).  In the non-ideal case we will use these exact minimizers of the ideal part of the free energy and evaluate the extra, non-ideal cost associated with them, thus forming an upper bound.  In reality an elastomer could, by adjusting the laminates, or by finding an unlikely continuous field, lower the energy by correlating distortion fields with the random field.  We discuss that energy reduction strategy briefly later.

\subsection{Ideal isotropic genesis polydomains}
The only completely solvable system is that of ideal isotropic genesis (iI). The energy function has no $\n_0$ dependence so it is not really spatially heterogenous at all. This is unsurprising since the isotropic cross-linking state appears to be completely homogeneous. It is the same model as is used for ideal monodomain samples, so it has been intensively studied. Its key property is that it is minimized locally by deformations of the form $\matr{\gamma}=\nabla\vec{y}=\matr{\ell}^{1/2}$, uniaxial stretches by $r^{1/3}$ along $\n$ which can be in any direction. These correspond to the spontaneous deformations that are caused by the transition to an aligned state. Since there are no preferred directions for alignment, one possible deformation field that allows every domain to be stretched uniaxially by $r^{1/3}$ is simply for the whole sample to stretch this much along the same axis, so the deformation gradient is $\nabla\vec{y}=\matr{\ell}^{1/2}$ at every point in the sample. This is a macroscopically aligned, monodomain, state with the director along the stretch axis and is a minimum energy state on the boundary of the quasi-convex hull. Clearly any direction could have been chosen for the stretch, so there is a degenerate set of equivalent low energy states each with a different deformation with respect to the cross-linking state, and deformations that map between these states must be soft.

\begin{figure}[htp]
\includegraphics[width=8cm]{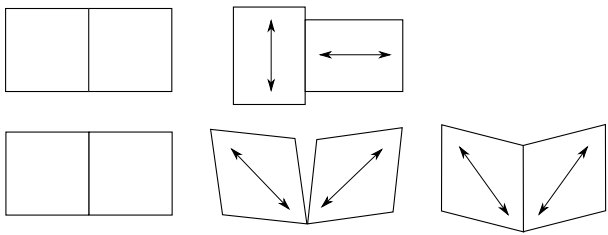}\caption{Top: Two uniaxial stretches with different axes applied either side of a boundary generally result in the material fracturing at the boundary. Bottom: If the boundary normal bisects the two axes then the boundary is stretched to the same degree under both deformations and material continuity can be restored by body rotating the two regions back together.}\label{twindiagram}
\end{figure}

Other macroscopic deformations will also be minimum energy states if they can be realized as the average of a compatible deformation fields that are uniaxial, volume-preserving stretches by $r^{1/3}$ at every point in the elastomer -- these macroscopic deformations lie on the interior of the quasi-convex hull. However, the condition of mechanical compatibility places a very strict condition on these deformation patterns. If two adjacent regions undergo different uniaxial stretches then this will generally lead to the material fracturing along the boundary between the regions. However, if the regions are separated by a plane boundary and the boundary bisects the angle between the two stretches, then the boundary will be stretched to the same degree by both deformations and material continuity can be restored by simply body rotating both regions. The fracturing and compatible  possibilities are illustrated in Fig.\ \ref{twindiagram}. This latter type of deformation pattern can be repeated many times to make a textured deformation that achieves a total deformation that is not simply a uniaxial stretch by $r^{1/3}$ whilst ensuring that the deformation at every point is such a stretch. The structured type of director and deformation pattern this leads to -- oscillations between regions of constant deformation and director separated by plane boundaries -- is characteristic of textured deformations that are driven by the requirement of mechanical compatibility between domains \cite{Mom, DeDo2002}.  The stripe-domains observed in monodomain samples \cite{kundler97, zubarevstripes} are another example of this type of behavior.

The full relaxation, $F^r(\matr{\lambda})$, of the ideal isotropic genesis energy has already been found \cite{DeDo2002} in the context of ideal monodomain elastomers. Remarkably, there are textured deformations that allow any macroscopic deformation with largest principal stretch less than $r^{1/3}$ and smallest principal stretch greater than $r^{-1/6}$ to be achieved whilst every point in the material undergoes a local uniaxial stretch of $r^{1/3}$. This class includes the deformation $\matr{\lambda}=\dm$, so the elastomer can accommodate its spontaneous deformation without changing shape macroscopically, and every uniaxial stretch up to $r^{1/3}$, so the elastomer can be stretched by any degree up to $r^{1/3}$ without the energy of the elastomer rising. After a uniaxial stretch of $r^{1/3}$ the elastomer is in a completely aligned monodomain state, so further stretching in the same direction cannot lead to further director rotation and the elastomer deforms without texture and is no longer soft.

The full relaxation result in \cite{DeDo2002} is that, if the principal stretches of $\matr{\lambda}$ are $f_1\le f_2\le f_3$,
\begin{equation}
\frac{2 F^{r}_{iI}(\matr{\lambda})}{\mu}=\begin{cases}
3 &\text{if $\matr{\lambda} \in K^{qc}$}\\
r^{1/3}(2/(r^{1/2}f_1)+f_1^2)&\text{if $\matr{\lambda} \in I$}\\
r^{1/3}( f_1^2+f_2^2+{f_3^2}/{r})&\text{if $\matr{\lambda} \in S$}\\
\infty &\text{else}
\end{cases}\label{relaxedenergy}
\end{equation}
where
\begin{align*}
I&=\{\matr{\lambda} \in \mathbb{M}^{3\times 3}:\frac{f_3}{f_2}<\sqrt{r},f_1<\frac{1}{r^{1/6}},\Det{\matr{\lambda}}=1\}\\
S&=\{\matr{\lambda} \in \mathbb{M}^{3\times 3}:\frac{f_3}{f_2}\ge\sqrt{r},\Det{\matr{\lambda}}=1\}\\
K^{qc}&= \{\matr{\lambda} \in \mathbb{M}^{3\times 3}:r^{-1/6}\le f_1 \le f_3 \le r^{1/3},\Det{\matr{\lambda}}=1\}.
\end{align*}
The ``else'' case only contains deformations with $\Det{\matr{\lambda}}\neq1$ that do not conserve volume. Textured deformations are required for macroscopic deformations in $I$ and $K^{qc}$ but not $S$. The set $K^{qc}$  is an eight dimensional set of zero energy deformations, so if an elastomer is at a point in the interior of  $K^{qc}$ all small volume preserving deformations of the elastomer are also in  $K^{qc}$ and do not cost energy to impose -- the the energy is liquid-like. In $S$ the energy depends on all three principal values of the deformation so the energy is solid like while in $I$ it depends on only the smallest principal value so the energy is intermediate between that of a solid and that of a liquid.  Physically, the set $S$ consists of stretches, $f_3$, sufficient that the director has completely aligned with this stretch direction, so the elastomer responds to further stretching in the same way as a conventional rubber, while the set  $K^{qc}$ consists of deformations that can be made soft by the formation of textured deformations of spontaneous deformations. The set $I$ contains  compression in one direction making the sample thinner than any  spontaneous deformation would. The director lies perpendicular to the compression axis but still forms textures in that plane allowing any macroscopic in-plane shapes bounded by $f_3/f_2 < \sqrt{r}$. The energy however only depends on the degree of compression.

In this work we are concerned with uniaxially stretching polydomains to induce the polydomain-monodomain transition. This corresponds to macroscopic deformations of the form $\matr{\lambda}=\rm{diag}(\lambda,1/\sqrt{\lambda},1/\sqrt{\lambda})$ for $\lambda\ge1$. Applying the above relaxation result, these deformations will be achieved at the energies
\begin{equation}
F_{iI}(\lambda)=\begin{cases}
\frac{3\mu}{2} & \text{if $1\le\lambda\le r^{1/3}$}\\
\frac{\mu r^{1/3}}{2}\left(\frac{2}{\lambda}+\frac{\lambda^2}{r}\right) & \text{if $\lambda\ge r^{1/3}$.}
\end{cases}\label{idealenergy}
\end{equation}
Differentiating this gives the engineering stress,
\begin{equation}
\sigma_{iI}(\lambda)=\frac{dF_{iI}}{d\lambda}=\begin{cases}
0 &\lambda \le r^{1/3}\\
\mu r^{1/3} (\lambda/r-1/\lambda^{2})&\lambda \ge r^{1/3}.
\end{cases}
\end{equation}
This simply means that ideal isotropic genesis polydomains will deform at zero stress but with textured deformations until they have been stretched by $r^{1/3}$, at which point they are completely aligned monodomains and respond to further deformation as a neo-Hookean solid. This result is quantitatively wrong, real isotropic genesis polydomains do not deform a absolutely zero stress, and when stress is removed they (macroscopically) return to their original configuration. This motivates the consideration of non-ideal theories. However, since the cross-linking state is almost isotropic, non-ideality must be very small so we expect several key features of the ideal model to persist. Extension will not occur at zero stress, but extensions up to $\lambda\le r^{1/3}$ will occur at energies $O(\alpha)\ll 1$. Furthermore, the deformation patterns and director patterns in the non-ideal case must still be very close in energy to those in the ideal case, so the observed patterns will still be characterized as textured deformations driven by elastic compatibility -- not disclination textures. Indeed, nematic disclinations never have zero-energy elastically compatible associated distortions \cite{disc}, so cannot be observed in isotropic genesis systems.

\subsection{Non-Ideal isotropic genesis polydomains}
Finding the full relaxation of the non-ideal polydomain energy is probably intractable. At the moment the relaxation is not known for the easier monodomain case (except in a thin film limit \cite{Contisemisoft}), and the polydomain result will depend to some extent on the exact form of $\n_0(\vec{x})$. However, we can put upper and lower bounds on the energy-strain curves. Developing an upper bound on the energy is straightforward. We simply use a textured test strain field from the ideal case and calculate its energy in the non ideal case. Since the relaxed energy function is a minimum over all strain fields, evaluating the energy at one example of a strain field is an upper bound on the energy.  In a sense this is a Taylor-like bound of uniform strain, but our uniform macroscopic strain is in fact composed of textures that allow deformation anywhere in the QCH.  We depart also from conventional Taylor bounds in that our bound is valid for large strains, up to $r^{1/3}$ which can be 100s\% for nematic elastomers.

If a point in the elastomer undergoes a uniaxial extension of magnitude $\gamma$ from the crosslinking state, at an angle $\theta$ to the preferred direction $\n_0$, then the energy of the deformation is
\begin{align}
F&=\min_{\n}\frac{\mu}{2}\Tr{\matr{\gamma}\cdot\matr{\gamma}^T\! \cdot{\matr{\ell}^{-1}}+\alpha r^{1/3} \matr{\gamma}\left(\dm-\n_0\n_0\right)\matr{\gamma}^T\n\n}\notag \\ &=\half\mu r^{1/3} \left(\frac{2}{\gamma}+\left(\frac{1}{r}+\alpha\right)\gamma^2-\alpha\gamma^2\cos^2{\theta}\right)\label{sintheta},\end{align}
where the minimization over $\n$ is achieved by taking $\n$ along the axis of $\matr{\gamma}$.
If the region of constant deformation is much larger than the individual domain size (region of given $\n_0$) then averaging over $\n_0$ gives:
\begin{align}
F=\half\mu r^{1/3} \left(\frac{2}{\gamma}+\left(\frac{1}{r}+\alpha\right)\gamma^2-\frac{\alpha\gamma^2}{3}\right),\end{align}
which is minimized at $\gamma_m^3=r/(1+2\alpha r/3)$ with a value
\begin{equation}
F=\frac{3\mu}{2}\left(1+\frac{2\alpha r}{3}\right)^{1/3}.
\end{equation}

This free energy density can be achieved if every point in the sample undergoes a uniaxial elongation of magnitude $\gamma_m$ in any direction. This situation is completely analogous to the situation in ideal elastomers in the isotropic configuration where the same (in this case minimal) energy can be reached by applying an elongation of magnitude $r^{1/3}$ in any direction. The DeSimone and Dolzmann texture result \cite{DeDo2002} shows that any uniaxial macroscopic deformation with magnitude less than $\gamma_m$ can be achieved by a texture of deformations in which each deformation is a uniaxial deformation by $\gamma_m$. This allows us to place an upper bound on the total energy of the sample after it has undergone a macroscopic uniaxial elongation by $\lambda$,
\begin{equation}
\frac{2 F^r(\lambda)}{\mu}\le\begin{cases} 3\left(1+\frac{2\alpha r}{3}\right)^{1/3} & \lambda^3\le r/(1+2\alpha r/3) \\
r^{1/3}\left(\frac{2}{\lambda}+\left(\frac{1}{r}+\frac{2\alpha}{3}\right)\lambda^2\right)& \lambda^3\ge r/(1+2\alpha r/3).
\end{cases}
\end{equation}
Although this upper bound has been calculated by using textures with regions of constant deformation that are very large compared to the length scales $\n_0$ varies on, the same result can be achieved with any size. This is because the averaging to $1/3$ of the $\cos^2{\theta}$ in eq.\ \eqref{sintheta} will still be true after averaging over many domains provided the axes of the domains are not correlated with the $\n_0$ field. Introducing such correlations would reduce the energy of the elastomer and would also determine length scales involved in the actual deformation and director fields.  An Imry-Ma style attack on this problem, but not involving textured test fields as here, is due to Terentjev and Fridrikh \cite{terentjevpoly}.  Domain size would be selected to take advantage of fluctuations in the random ordering field from crosslinking.  We return to this problem elsewhere.

The ideal system provides a very simple lower bound on the energy, eq.\ \eqref{idealenergy}, since the non-ideal term is never negative. We can improve on this bound by using the Sachs limit of stress uniform through the sample. This provides a lower bound because it neglects the requirement of compatibility of deformations ($\matr{\gamma}=\nabla\vec{y}$). We calculate this bound numerically by minimizing $F_{nI}\left(\matr{\gamma},\n,\n_0\right)-\sigma\gamma_{xx}$ across all $\matr{\gamma}$ and $\n$ at fixed $\sigma$ for a given domain ($\n_0$) to find the optimal deformation $\matr{\gamma}_m$ and director orientation $\n_m$ of the domain at the stress $\sigma$. The energy and extension of the whole sample are then found by averaging $F_{nI}\left(\matr{\gamma}_m,\n_m,\n_0\right)$ and $\left(\matr{\gamma}_m\right)_{xx}$ across all domain orientations.

Although we have calculated the full Sachs bound numerically, we can understand its behavior at small extension analytically. At zero stress every domain is free to undergo an energy minimizing spontaneous deformation - anything of the form $\matr{\gamma}=\matr{R}\cdot\matr{\ell_0}^{1/2}$, where $\matr{R}$ is a rotation. The $xx$ component of this is simply $\vec{x}\cdot\matr{R}\cdot\matr{\ell_0}^{1/2}\cdot\vec{x}$, which is to say it is the component of $\matr{\gamma}\cdot\vec{x}$ that is parallel to $\vec{x}$. For each domain a rotation $\matr{R}$ can be chosen such that $\matr{\lambda}_{xx}$ lies anywhere between $0$ and $|\matr{\ell_0}^{1/2}\cdot\vec{x}|$. Since we are studying extension we are not concerned with $\lambda_{xx}\le1$, but this means that any $\lambda_{xx}$ between $1$ and  $\left<|\matr{\ell_0}^{1/2}\cdot\vec{x}|\right>$ can, in the Sach's limit, be achieved without stress. Therefore the Sach's limit on the energy is simply $3\mu/2$ for $1\le\lambda_{xx}\le\left<|\matr{\ell_0}^{1/2}\cdot\vec{x}|\right>$. It is straightforward to calculate this average giving
\begin{equation}
\left<|\matr{\ell_0}^{1/2}\cdot\vec{x}|\right>=\frac{r^{-1/6}}{2}\left( \sqrt{r}+\frac{\sinh^{-1}{\sqrt{r-1}}}{\sqrt{r-1}} \right).
\end{equation}
This threshold tends to $1+O(\epsilon^2)$ for $r=1+\epsilon$, so for small anisotropies it is insignificant, but for large $r$ it tends to $r^{1/3}/2$.

\begin{figure}
\includegraphics[width=8cm]{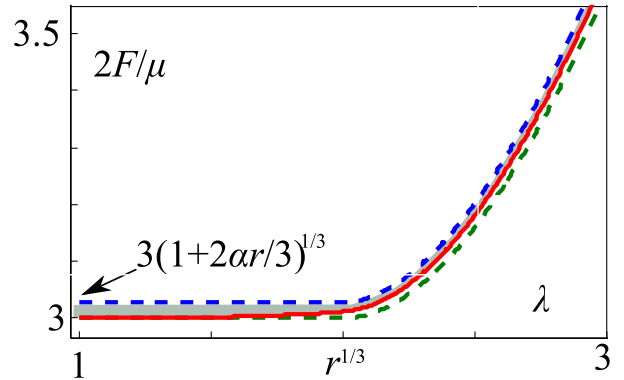}
\caption{Bounds on the free energy density of an isotropic genesis polydomain nematic elastomer as a function of strain, plotted in units of $\mu r^{1/3}/2$ with $r=8$ and $\alpha=0.005$. Upper curve (dashed, blue): upper bound from a test strain field. Middle curve (smooth, red): Sachs lower bound. Lower curve (dashed, green): ideal result, lowest bound. }\label{energybounds}
\end{figure}

The energy plot, Fig.\ \ref{energybounds}, shows that the bounds constrain the energy very tightly: the bounds reproduce the exact ideal result if $\alpha=0$, and $\alpha$ is expected to be very small for isotropic genesis. 
The Sachs free energy is in effect plotted parametrically since one sets the stress and obtains the free energy and strain by minimisation and averaging.
The bounds show that stretching the elastomer by $\sim r^{1/3}$ cannot require an increase in the energy density of more than $3\mu/2((1+2\alpha r/3)^{1/3}-1)\approx \mu\alpha r/3$. This means that although the extensions up to $r^{1/3}$ can take place at finite stress, the stress cannot be higher than $\partial F/\partial \lambda \sim \mu\alpha r/[3(r^{1/3}-1)]$, which will be a very small number since $\alpha$ is small. As in the ideal case, extensions larger than $\sim r^{1/3}$ behave in a neo-Hookean manner.

We can calculate bounds on the stress-strain curve for these extensions by using the requirement, which applies to all one dimensional elastic energies, that the relaxed energy curve be convex, meaning the stress curve is monotonic for $\lambda\ge1$. This means that at a given extension $\lambda_{xx}$ not only must the energy function lie between the two bounds, but the gradient of the energy function must not be so great that, if extrapolated forwards as a straight line, the energy curve intersects the upper-bound (used above for estimating the approximate maximal stress), or so little that if extrapolated backwards it does the same. This bounds the gradient of the energy and hence the stress at each extension.

\begin{figure}
\includegraphics[width=8cm]{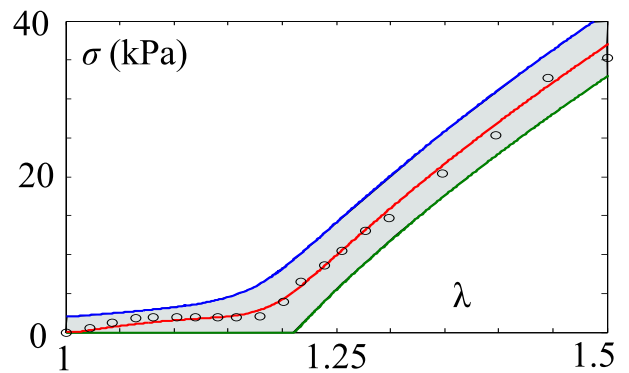}\caption{Upper and lower bounds on the stress-strain curve for a non-ideal isotropic genesis polydomain. Material Parameters $r=1.65$ and $\alpha=0.01$ and $\mu=33000$. The middle curve is the derivative of the Sachs lower bound on the energy, which provides as estimate of the stress-strain curve, while the upper and lower curves are the bounds derived by requiring the energy function be convex.  The circles are stress-strain data for a real sample. We thank K.\ Urayama for permission to reproduce this data.}\label{stressstrain}
\end{figure}

 A plot of the two bounds on the stress strain curve is shown in Fig.\ \ref{stressstrain}, which shows that the very soft stress plateau for $\lambda\le r^{1/3}$ does indeed survive the introduction of semi-softness.
The experimental results of Urayama \textit{et al} for isotropic genesis polydomain stress \cite{Uryamapoly} are also shown and display pronounced softness -- clearly polydomains can deform softly and the requirement of compatibility between domains does not appreciably harden response.

Random field
models  of polydomain nematic
elastomers \cite{terentjevrandom,terentjevpoly,Uchida:99a,Uchida:99b,Uchida:00} address the question of how polydomain elastomers form, and how the domain structure varies with temperature and stress.  Uchida, in  2-D numerical simulations of domains, found disclinated structures. He also
concluded that nematic elastomers cross-linked in the isotropic
state should display soft-behavior while those cross-linked in the
nematic state would not, using a combination of infinitesimal
analysis and finite strain simulation, both also in 2-D.
 Our finite-strain analysis  has been very
different: we employ multi-scale test textures giving
identically zero upper and lower bounds on elastic energies (for
ideal isotropic genesis elastomers) and low energy response in the
presence of disorder. In common with Uchida, we obtain hard
response for nematic genesis elastomers:

\subsection{Nematic genesis polydomains}
We expect the coefficient of non-ideality, $\alpha$, to be much higher for nematic genesis polydomains because the cross-linking state is not isotropic and distinguishes the direction of the nematic director, $\n_0$. It will probably have a similar magnitude to that observed in monodomain elastomers where $\alpha\sim 0.1$. However, although non-ideality will be larger, it is conceptually less important because already the ideal part of the energy will contribute significantly to the stress. This is because even an ideal nematic genesis polydomain is not expected to have any soft modes. Soft modes are generated by symmetry breaking spontaneous distortions from an isotropic reference state. Locally an individual domain does have an isotropic reference configuration, which is reached by applying the  inverse spontaneous deformation $\matr{\gamma}=\matr{\ell_0}^{-1/2}$, but the spatially-dependent deformation  $\matr{\gamma}(\vec{x})=\matr{\ell_0}^{-1/2}(\vec{x})$ will not be mechanically compatible, so it is impossible to apply a deformation that places the whole sample in the isotropic reference state simultaneously. The reason for this large difference between nematic and isotropic genesis polydomains is that there are very few deformation patterns of the form  $\matr{\lambda}(\vec{x})=\matr{\ell_0}^{-1/2}(\vec{x})$ that are compatible deformations, and no nematic disclination patterns are compatible deformations. An isotropic genesis polydomain is forced to undergo such a deformation on cooling to the nematic after cross-linking, so it must choose one of the deformation patterns that is mechanically compatible and these, as we have seen in the previous section, allow for macroscopic movement across the QCH. In contrast the nematic-genesis polydomain undergoes its ``spontaneous-distortion'' in the melt where there are no conditions of compatibility because adjacent regions can flow past each other, a deformation   which would result in a fracture in a solid network. One consequence of this is that when a nematic genesis polydomain is heated to what would be its  ``isotropic'' state, it is unable to undergo its energy minimizing contraction $\matr{\gamma}(\vec{x})=\matr{\ell_0}^{-1/2}(\vec{x})$ everywhere because this is not a compatible deformation. This will result in the high temperature state being internally stressed, and may lead to elevation of the transition temperature.

We can bound the ideal nematic genesis free energy in the same way we bounded the isotropic genesis non-ideal free energy. An upper bound is provided by a test strain field. The simplest test field would be a constant strain throughout the sample, giving a Taylor bound.  However, we can find a tighter upper bound by applying the same strain to each domain, but allowing each domain to form textured deformations that average to the strain imposed on it; this is what we termed in the previous section a ``Taylor-like bound".  Consequently, if the strain imposed macroscopically is $\matr{\lambda}$, the free energy density of each domain will be $F^r_{Ii}(\matr{\lambda}\cdot\matr{\ell}^{1/2})$, where $F^r_{Ii}(\matr{\lambda})$ is the relaxed energy for an ideal monodomain in \cite{DeDo2002} given by eq.\ \eqref{relaxedenergy} and is attained by texturing (by laminates). The factor of $\matr{\ell}^{1/2}$ in the argument of $F^r_{Ii}$ is appropriate because the function  $F^r_{Ii}(\matr{\lambda})$ is written in terms of deformations from the isotropic reference state, whereas the domains in the nematic genesis polydomain are already in the elongated nematic state. The deformation $\matr{\ell}^{1/2}$ is the deformation the isotropic state would have to undergo to reach the nematic state that the domain was cross-linked in, and the component $\matr{\lambda}$, of the compound deformation  in the argument of $F^r_{Ii}$, is the deformation from this nematic state. A lower bound can be found using the Sachs constant stress limit. However, since the elastomer is ideal, the individual domains can deform softly. This means that, in the constant stress limit where compatibility of the deformations between the different domains is not required, the elastomer can deform completely softly. We can calculate the end of the soft plateau in the Sachs bound analytically in the same way we did for the isotropic-genesis case.  Since one considers a spontaneous contraction  $\matr{\ell_0}^{-1/2}$ in going from the nematic state to an isotropic reference state, followed by a spontaneous elongation $r^{1/3}$ in elongating to the nematic state along $\n$, the end softness in the Sachs bound will occur at $\left<|r^{1/3}\matr{\ell_0}^{-1/2}\vec{x}|\right>$ which evaluates to
\begin{equation}
\left<|r^{1/3}\matr{\ell_0}^{-1/2}\vec{x}|\right>=\frac{1}{2}\left(1+\frac{r \tan^{-1}{\sqrt{r-1}}}{\sqrt{r-1}}\right).
\end{equation}
This limits to $1+\epsilon/4$ for small anisotropy $r=1+\epsilon$, and to $\sqrt{r}\pi/4$ for large $r$.

\begin{figure}
\includegraphics[width=6cm]{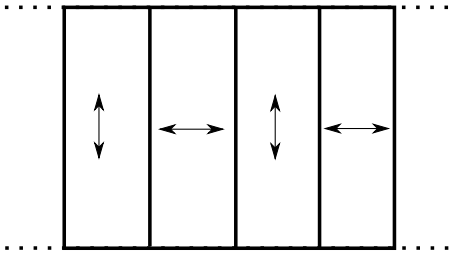}\caption{An example of a possible director pattern at cross-linking in a nematic genesis polydomain that certainly doesn't have any macroscopic soft modes. This can be seen by considering a line element along the boundary in the plane of the diagram - all soft modes of the vertical stripes require this element to contract while all soft modes of the horizontal stripes require it to extend or not deform, so there are no macroscopic soft deformations.}\label{fig:nemstripehard}
\end{figure}

\begin{figure}
\includegraphics[width=8cm]{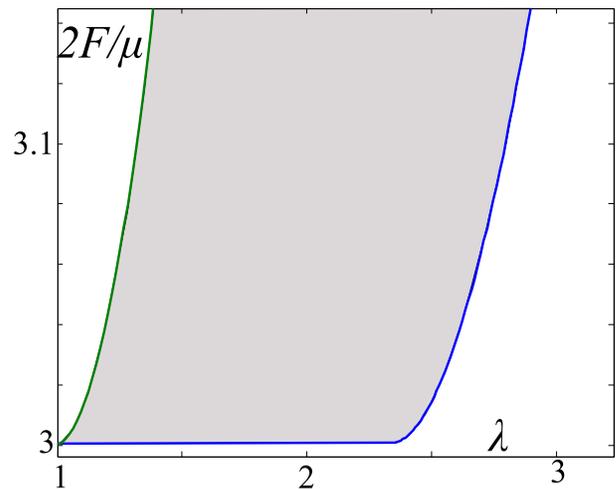}\caption{Bounds on the free energy density of an ideal nematic genesis polydomain with $r=8$.  Upper curve (green) --- upper bound using a test strain field. Lower curve (blue) --- lower bound using a constant stress. There is a very large gap between the two bounds.  Note that the ideal system displays hard elastic response for purely geometrical reasons stemming from compatibility requirements.}\label{idealnemticbounds}
\end{figure}

\begin{figure}
\includegraphics[width=8cm]{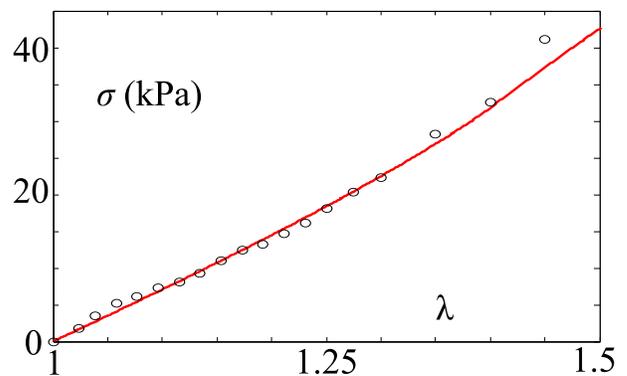}\caption{Estimate of stress vs strain for an ideal nematic genesis polydomain elastomer with $r=1.65$ and $\mu=37000$, obtained by differentiating the upper bound on the energy. The circles are stress-strain data for a real sample, the exact analogue of the sample used in fig.\ \ref{stressstrain} but cross-linked in the nematic polydomain state.  We thank K.\ Urayama for permission to reproduce this data. We note that the fitting parameter $r$ is the same but a slightly higher value of $\mu$ is needed. This probably reflects the fact that in reality nematic genesis polydomains have a significant non-ideal term ($\alpha\sim0.1$) which will slightly harden the system.}\label{nemgentaylorstress}
\end{figure}

The two bounds on the energy (calculated numerically) are both plotted in Fig. \ref{idealnemticbounds}. These bounds on the energy are not very good --- there is a large gap between them. The Sach's limit displays complete ordering of the elastomer at zero stress while the test strain field shows hard elasticity and finite modulus ($\sim\mu$) at all extensions. This is because nematic genesis polydomains are strongly heterogeneous materials, and good methods for finding stress strain curves for such materials have yet to be developed at large deformations that are required here. Taylor (affine) and Sach's (constant stress) bounds on heterogeneous materials are independent of the domain structure of the material, meaning the Taylor bound gives an upper bound on the energy of the hardest possible domain structure, and Sach's gives a lower bound on the energy for the softest possible domain structure. Unfortunately there exist vanishingly unlikely domain structures which are indeed completely soft, namely the textured-deformation domain structures realized by the isotropic genesis polydomains. There also exist domain structures which certainly have no macroscopic soft modes - fig.\  \ref{fig:nemstripehard}.
Since the domain structures in nematic polydomains will certainly not be of the compatible double laminate type --- they will be disclination textures which are not mechanically compatible --- the lower bound is of little physical significance. Therefore, we estimate the stress strain relation for these polydomains by simply taking the derivative of the upper bound on the energy. The resulting stress-strain curve, shown in Fig.\ \ref{nemgentaylorstress}, matches the experimentally observed completely hard behavior.

The inclusion of non-ideality in the nematic genesis model will elevate the Sach's limit to a plateau of height $\propto\alpha$ but will not alleviate the fundamental difference between the soft elasticity seen in the Sachs and the hard elasticity seen in the Taylor-like limit.

\section{Smectic Polydomains}
Smectic liquid crystal phases are phases in which the rods not only have orientational order but are also layered. Liquid crystal elastomers can exhibit smectic ordering \cite{finkelmann1981}, and monodomains with both SmA ordering \cite{fischer1995}  (in which the liquid crystal director is parallel to the layer normal) and SmC ordering \cite{benne1994} (in which the liquid crystal director makes a constant angle $\theta$ with the layer normal) have been synthesized. SmC phases in which all the rods have the same chirality (SmC* phases) are particularly interesting because they have electrical polarizations along the cross product of the layer normal and the director. The introduction of these phases significantly increases the total number of polydomains that can be considered since there are now four distinct states --- isotropic, nematic, SmA and SmC --- and polydomains can be made that have been cross-linked in any one of these states then cooled or heated to any other of the states.

In strongly coupled elastomers, smectic behavior is usually modeled by assuming that the layers deform affinely under deformations (so that a layer normal $\vec{k}$ becomes $\matr{\gamma}^{-T}\cdot \vec{k}$ after a deformation $\matr{\gamma}$ has been applied) and then adding terms to the underlying nematic free energy that penalize changing the inter-layer spacing and rotating the director away from its preferred angle with the layer normal. This means that although SmA elastomers have the same cylindrical symmetry as nematics they do not have any soft elastic modes since the layers deform affinely (they cannot translate or rotate relative to the rubber matrix) and rotation of the director away from the layer normal costs energy. SmC elastomers could still exhibit soft modes since their director can rotate in a cone around the layer normal without changing the inter-layer spacing or deviating from the preferred tilt angle.

\noindent (i) \textit{Isotropic Genesis}: The modeling assumptions outlined above are well established for monodomain smectic elastomers but they seem rather strange for polydomain elastomers crosslinked in the isotropic state. In particular, the assumption that the layers move affinely seems appropriate if the layers have been embedded into the elastomer at crosslinking.  But if, for isotropic genesis, they have appeared after crosslinking as the result of a symmetry-breaking isotropic-SmA transition, then they could equally well have formed in any other direction so one would expect them to be able to rotate through the sample. Indeed, since the isotropic-SmA transition will be accompanied by a spontaneous deformation that is a stretch along the director and hence also the layer normal, a deformation that returns the elastomer to its isotropic configuration and then stretches it by the same amount in a different direction must, on symmetry grounds, be soft and must cause the layers to rotate. With this view, the isotropic cross-linked SmA polydomains are no different to the isotropic cross-linked nematic polydomains since the spontaneous deformations at the transitions are of exactly the same form and break the same symmetry.  Accordingly we expect exactly the same macroscopic soft elastic response. SmC polydomains cross-linked in the isotropic state will also behave in the same way. However, although this simple symmetry argument cannot be circumvented in equilibrium, it is possible that non-equilibrium kinetic effects prevent the elastomer from realizing this soft elasticity on experimentally accessible time scales and that the layers, though really free to rotate through the sample, are effectively frozen in at the transition to the layered state. This freezing would result in a complete hardening of SmA elastomers since, while the layers are deforming affinely, there are no local soft modes.  Equally, affine layer deformations imposed by the freezing-in of layers would make the SmC polydomain much like the nematic genesis nematic polydomain case since it would possess local soft modes but there would be no compatibility between the soft modes of adjacent domains so these cannot be used to make macroscopic soft modes.

\noindent (ii) \textit{SmA or SmC Genesis}: Most of the other routes to smectic polydomains will result in macroscopically hard elasticity, for example cross-linking directly in a SmA state will lead to a polydomain with no local soft modes, cross-linking in a SmC polydomain state will result in a system analogous to the nematic cross linking nematic polydomain.

\subsection{Soft polydomain smectic elasticity}\label{subsect:softSmectic}
\begin{figure}[htp]
\includegraphics[width=8cm]{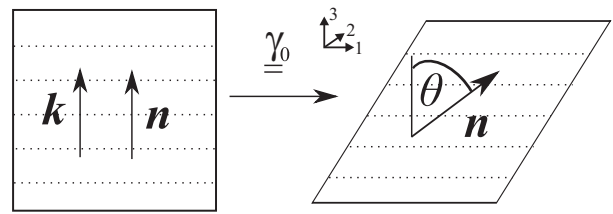}
\caption{A SmA elastomer (left) with director $\vec{n}$ aligned along the layer normal $\vec{k}$ cools and undergoes a deformation $\matr{\gamma_0}$ to form a SmC elastomer (right) in which the director forms an angle $\theta$ with the layer normal. The deformation typically includes a contraction along $\vec{k}$ since the rods have tilted which reduces the inter-layer spacing. }\label{smcdiagram}
\end{figure}

To recover macroscopically soft elasticity, we need the elastomer to break a symmetry at a transition after cross-linking in such a way that if it had happened to break it in the same way in every domain a monodomain would have formed. One interesting system that has this property is a SmA monodomain that is cooled without any external influences into a SmC polydomain. Crosslinking in the SmA monodomain state guarantees that the smectic layers are permanently embedded in the elastomer and will subsequently deform affinely and that, since the layers are embedded as a monodomain, after cooling the system can access fully aligned SmC monodomain states without having to rotate the layers through the elastomer. To discuss this case will take the SmA cross-linking state as the reference state and use a set of axes such that the SmA layer normal $\vec{k}=(0,0,1)$ while $(0,1,0)$ and $(1,0,0)$ are perpendicular vectors in the SmA layer plane, see Fig.\ \ref{smcdiagram}.
The local spontaneous deformations at the SmA-SmC transition if the director tilts in the $\vec{k}-(1,0,0)$ plane will be
\begin{equation}
\matr{\gamma_0}=\left( \begin{array}{ccc}
\gamma_{11} & 0 & \gamma_{13} \\
0 & 1/\gamma_{11}\gamma_{33} & 0 \\
0 & 0 & \gamma_{33} \end{array} \right)\label{eq:spont-shear}
\end{equation}
where all the components of this deformation have fixed values determined by the microscopic details of the elastomer. This transition is illustrated in Fig.\ \ref{smcdiagram}. Elastomers have been synthesized with values of $\gamma_{13}$ as large as 0.4 \cite{hiraoka}.

In the SmA phase there is nothing to distinguish any direction in the $(0,1,0)-(1,0,0)$ plane, so the above deformation would also have been soft if it had been applied at any other angle in the $(0,1,0)-(1,0,0)$ plane, so the full set of soft deformations, $K^0_{SmC}$, is all deformations that can be written in the form $\matr{R}\cdot\matr{\gamma_0}\cdot\matr{R}_\vec{k}$, where $\matr{R}$ is a rotation and $\matr{R}_\vec{k}$ is a rotation about $\vec{k}$. The full set of deformations that can be made soft by constructing textures out of such deformations is known to be \cite{contiadams}
\begin{align}
K^{qc}_{SmC}= &\{\matr{\lambda} \in \mathbb{M}^{3\times 3}: \Det{\matr{\lambda}}=1,| \matr{\lambda}\cdot\vec{k}|^2\le\gamma_{13}^2+\gamma_{33}^2, \notag \\&| \matr{\lambda}^{-T}\cdot\vec{k}|^2\le1/\gamma_{33}^2, f_1(\matr{\lambda}\cdot\matr{\lambda_i})\ge1 \},
\end{align}
where the function $f_1$ returns the smallest principal value of its argument and the deformation $\matr{\lambda}_i=\rm{diag}(\gamma_{11}\gamma_{33},  \gamma_{11}\gamma_{33},  \rho)$ where
\begin{equation}
\rho=\frac{1-\gamma_{11}^4\gamma_{33}^2}{\gamma_{13}^2+\gamma_{33}^2-\gamma_{13}^4\gamma_{33}^4}.
\end{equation}
The set $K^{qc}_{SmC}$ is fairly complicated, but the four conditions in it each have a simple interpretation. The determinant condition $\Det{\matr{\lambda}}=1$ requires that the elastomer have the same volume after deformation as it did in the SmA state.  A line element in the $\vec{k}$ direction of the SmA is stretched by a factor of $\sqrt{\gamma_{13}^2+\gamma_{33}^2}$ by any of the local spontaneous deformations in $K^0_{SmC}$, so the longest a line element in the $\vec{k}$ direction can be after deformation is $\sqrt{\gamma_{13}^2+\gamma_{33}^2}$ (which occurs when every local spontaneous deformation is the same), imposing the first inequality. The second inequality can be understood in an analogous  way --- a vector area along $\vec{k}$ has its area increased by a factor of $1/\gamma_{33}$ by any of the local spontaneous deformations but its direction may be rotated. The average of all these vector areas, $\matr{\lambda}^{-T}\cdot\vec{k}$, cannot be larger than the sum of the areas that make it up so $| \matr{\lambda}^{-T}\cdot\vec{k}|^2\le1/\gamma_{33}^2.$ The final inequality is the analog of $f_1(\matr{\lambda})\ge r^{-1/6}$ in the ideal nematic case, requiring that the elastomer cannot be compressed so much in any direction that it is thinner than the natural width of the thinnest direction of the underlying chain distribution.

The set $K^{qc}_{SmC}$ is much richer than its ideal nematic counterpart, in particular uniaxial deformations of the form $\matr{\lambda}=\rm{diag}(1/\sqrt{\lambda},1/\sqrt{\lambda},\lambda)$, which are stretches by $\lambda$ along the original layer normal, are in $K^{qc}_{SmC}$ provided that $\gamma_{33}^2\le\lambda^2\le\gamma_{33}^2+\gamma_{13}^2$. This means that there is a whole set of textured SmC polydomain states whose deformation with respect to the parent SmA state is a simple stretch along the SmA layer normal. All of these SmC polydomain states have the same macroscopic cylindrical symmetry as the SmA state. When the SmA sample is cooled to the SmC state it could form any of these states, so the resting configuration of a SmC polydomain may have a uniaxial stretch relative to the SmA state. This was not the case in the isotropic genesis nematic polydomains because there was only one textured state with the full isotropic symmetry of the cross-linking state.

In the ideal SmC case our inability to uniquely identify one polydomain state with cylindrical symmetry makes no difference to the analysis at all since every state in $K^{qc}_{SmC}$ is an energy minimizing state and deformations that move the polydomain between them are perfectly soft. The addition of a non ideal term, which must be small since it breaks both the homogeneity and symmetry of the cross-linking state by energetically favoring a single director orientation, will have a very similar effect to its addition in the nematic case. Non-ideality will break the complete energy degeneracy of $K^{qc}_{SmC}$ placing some states (those near the boundary where these is less freedom to choose between different textures to minimize the non-ideal energy) slightly higher in energy so that small but finite stresses are needed to allow the elastomer to explore the complete set. The lack of a single unique state with cylindrical symmetry means that we can not be sure what the energy minimizing state is, indeed which state it is will depend on the precise functional and spatial form of the non-ideal term included, so it may depend on the chemical nature of the elastomer.

\subsection{Electromechanical switching of soft polydomain SmC* elastomers}\label{subsect:soft-electromechanical}
Interest in SmC elastomers is mostly driven by their potential for electrical actuation. Chiral SmC* liquids  exhibit an (improper) ferro-electric polarization along the cross product of their director and layer normal \cite{Meyer:75}.  Being chiral there is a twist of the tilt direction (and hence of the polarization) about $\vec{k}$ on advancing along the layer normal direction.  There is accordingly no macroscopic electrical polarisation unless the twist is undone by external fields or by boundary effects \cite{Clark_Lagerwall:80}. On crosslinking in the SmA state and cooling to the SmC* state with domains, this twist is largely suppressed.  Twist of $\n$ about $\vec{k}$ means the $\gamma_{13}$ spontaneous shear direction in eqn.~(\ref{eq:spont-shear}) rotates from layer to layer, giving rise to elastic incompatibility.  The effect of such incompatibility on texture formation  has been discussed in detail in connection with mechanical switching of SmC monodomains \cite{Adams_switch:09}.  Without twist, domains accordingly develop a net polarisation  so that, when an electric field is applied, energy is minimized by domain reorientation so that polarization is parallel to the applied field. Our inability to specify which point deep in the interior of  $K^{qc}_{SmC}$ is the lowest energy resting state of SmA monodomain genesis SmC* polydomains does not prevent us from analyzing their electrical actuation since the most extreme actuation is achieved by making the elastomer traverse the whole set $K^{qc}_{SmC}$ (from boundary to boundary). If an electrical field is applied in the $(0,1,0)$ direction to such a SmC polydomain then this will cause it to form a state with its polarization vector uniformly in the $(0,1,0)$ direction which it can do by forming a monodomain with its director and layer normal both in the $(1,0,0)-(0,0,1)$ plane. The deformation of this monodomain with respect to the parent SmA state, $\matr{\gamma}_1$, is of the form of eqn~(\ref{eq:spont-shear}).
%\begin{equation}
%\matr{\gamma_1}=\left( \begin{array}{ccc}
%\gamma_{11} & 0 & \gamma_{13} \\
%0 & 1/\gamma_{11}\gamma_{33} & 0 \\
%0 & 0 & \gamma_{33} \end{array} \right).
%\end{equation}
If the electric field is then reversed the elastomer will flip into the opposite state which still has the director and layer normal both in the $(1,0,0)-(0,0,1)$ plane but with the director on the other side of the layer normal so that their cross-product (and hence the polarization) is reversed. This state has a deformation  with respect to the SmA 
\begin{equation}
\matr{\gamma}_2=\left( \begin{array}{ccc}
\gamma_{11} & 0 & -\gamma_{13} \\
0 & 1/\gamma_{11}\gamma_{33} & 0 \\
0 & 0 & \gamma_{33} \end{array} \right).
\end{equation}
The full deformation undergone be the elastomer when the electric field is reversed is $\matr{\lambda}={\matr{\gamma}_2}\cdot \matr{\gamma}_1^{-1}$, giving
\begin{equation}
\matr{\lambda}=\left( \begin{array}{ccc}
1 & 0 & -2\gamma_{13}/\gamma_{33} \\
0 & 1 & 0 \\
0 & 0 & 1 \end{array} \right),
\end{equation}
simply a reversal of the spontaneous shear.  Since the discrepancy in energy between different states in $K^{qc}_{SmC}$ is driven entirely by the addition of non-ideality which expected to be small, the electric fields required to perform this very large actuation will also be small --- smaller than the fields used to perform similar actuations on SmC* monodomain samples which have large non-ideal fields cross-linked into them to make them form monodomains. This suggests that SmA monodomain genesis SmC* polydomains are probably better candidates for electrical actuation than their monodomain counterparts.

\section{Conclusions}
There is a fundamental difference between those polydomain liquid crystal elastomers cross-linked in a high symmetry state then cooled to a low symmetry state and those crosslinked directly in the low symmetry state. The former will be extremely soft macroscopically while the latter will be mechanically hard. We have analyzed two completely soft examples, nematic polydomains cross-linked in the isotropic state and SmC polydomains cross-linked in the SmA monodomain state and one hard example - nematic polydomains cross-linked in the nematic state. This distinction between soft and hard polydomains has not previously been appreciated, but very recent experiments confirm that it is correct \cite{Uryamapoly}. The recognition of softness in some polydomain systems makes the fabrication of useful LCE soft actuators more likely since polydomains are much easier to synthesize and are not limited to thin film geometries.  Our results suggest that a SmC* polydomain cross-linked in a SmA monodomain state would be a good choice for low field electrical actuation.

LCE's are very analogous to martensitic metals since both systems exhibit symmetry-breaking transitions coupled to deformations. In the martensitic case the symmetries that are broken are discrete whereas in LCE's they are continuous. However, drawing analogies between LCE polydomains and martensite polycrystals is quite subtle. The soft polydomain LCE's with homogeneous high temperature cross-linking states are analogous to single crystal martensite systems. However, single martensite crystals are difficult to prepare because, even if they are prepared in the high symmetry state, they are not isotopic so the crystal can form with different orientations at different points in space. This is in marked contrast to isotropic genesis LCE polydomains where, because the cross-linking state is completely isotropic, cross-linking in a spatially homogeneous state is trivial. There is no satisfactory martensite analog of cross-linking in the low symmetry polydomain state since martensite poly crystals are formed in a poly crystalline high symmetry state, so they access a stress-free high symmetry state, whereas the low symmetry cross-linked elastomers cannot. There is one, albeit rather contrived, LCE system, not analyzed here, that is directly analogous to a martensite polycrystal --- an elastomer cross-linked in the SmA polydomain state then cooled to a SmC polydomain state. Such an elastomer would, on heating, return to a stress-free SmA polydomain state, and has local soft modes in the SmC state generated by the symmetry breaking SmA-SmC transition.

\bibliography{polylong}

\end{document}